\input mates.sty
\input textos.sty
\input formatos.sty
\font\twelverm=cmr12
\input epsf

\titpapb{Adel, Yndur\'ain}{High energy deep inelastic scattering}{1}

\rightline{FTUAM 96-44}
\rightline{ \datestamp}
\vskip1.5cm
\centerline{\twelverm  HIGH ENERGY  PHOTON DEEP INELASTIC SCATTERING AT   
SMALL AND LARGE $Q^2$}
\vskip.2cm
\centerline{\twelverm WITH SOFT PLUS HARD POMERON}
\vskip1.5cm
\centerline{\bf K. Adel,}
\vskip.1cm
\centerline{and}
\vskip.1cm
\centerline{{\bf  F. J. Yndur\'ain}\footnote*{\petit e-mail: fjy@delta.ft.uam.es}}
\vskip.3cm
\centerline{\sl Departamento de F\'\i sica Te\'orica, C-XI}
\centerline{\sl Universidad Aut\'onoma de Madrid}
\centerline{Canto Blanco, 28049-Madrid}
\vskip2.cm
\setbox0\vbox{\hsize110mm
\noindent{\bf ABSTRACT}.
\vskip.5cm
{\petit We show how the sum of a hard singularity, $F_{2H}(x,Q^2_0)\sim x^{-\lambda}$ and 
a soft Pomeron $F_{2P}(x,Q^2_0)\sim {\rm Const.}$ for the singlet piece 
of the structure function $F_{2S}=F_{2H}+F_{2P}$ for $Q_0^2\sim \;{\rm a\; few}\;\gev^2$, 
 plus a saturating expression for the strong coupling,
 $\tilde{\alpha}_s(Q^2)=4\pi/\beta_0\log[(Q^2+\Lambdav^2)/\Lambdav^2]$ 
give an excellent description of experiment {\it i}) For small $Q^2$, 
$0\lsim Q^2\leq 8.5\,\gev^2$, and {\it ii}) For large $Q^2$, $10\lsim Q^2\leq 1 500\,\gev^2$ if 
evolved with QCD. The $x$ range is $6\times10^{-6}\lsim x \lsim 0.04$. The 
description for low $Q^2$ implies self-consistent values for the parameters 
in the exponents of $x$ both for singlet 
and nonsinglet. One has to have $\alpha_{\rho}(0)=0.48$ and $\lambda=0.470$ [$\alpha_P(0)=1.470$], 
in uncanny agreement with other determinations of these parameters, and in particular
 the results of the large $Q^2$ fits. The fit to data is so good that we may look for 
signals of a ``triple Pomeron" vertex, for which some evidence is found.}}
\centerline{\box0}
\vfill\eject
\noindent\S{\bf 1. Introduction}. The HERA collaborations$^{[1,2,3]}$ have
 produced in the last years
 data with increasing precision for the $ep$ deep inelastic scattering (DIS) structure 
function $F_2(x,Q^2)$ at small $x$, and in a wide range 
of momenta, from $Q^2\sim 0$ (including real photoproduction at $Q^2=0$)
 to thousands of GeV$^2$. This allows us to study in detail not only 
the low $x$ behaviour at momenta sufficiently large\fonote{We will consider 
the region $Q^2\gsim 10\,\gev^2$ to be that where perturbative QCD is 
to be trusted. This is because, at $x\rightarrow 0$,
 the NLO corrections are {\it very} large.} 
that perturbative QCD be applicable, which was done in ref. 5
 in a full NLO (next to leading order) calculation, but also  
enables us to address the important issue of the connection between 
 the perturbative regime ($Q^2>10\,\gev^2$, say) and the region
 ($Q^2<10\,\gev^2$) 
where nonperturbative effects are determinant.

In ref. 5, where we send for more details, we studied fits of the large 
momentum region with the {\it hard singularity} hypothesis, i.e., 
under the assumption that the singlet component of $F_2$, $F_S$, behaves as 
a negative power of $x$ as $x\rightarrow 0$ so that, as discussed first 
in ref. 6, QCD implies the detailed behaviour
$$F_S(x,Q^2)\simeqsub_{x\rightarrow 0}B_S[\alpha_s(Q^2)]^{d_+(1+\lambda)}x^{-\lambda}
\eqno (1)$$
to leading order (LO). The {\it soft Pomeron dominance} hypothesis was also considered. Here 
one assumes$^{[7]}$ that at a given $Q^2_0\sim\,1\,\gev^2$ one has a standard 
Pomeron (constant) behaviour, and then evolves with QCD so that, to LO,
$$\eqalign{F_S\simeq
 \frac{c_0}{\xi}\left[ \frac{9\xi \tau}{4\pi^2(33-2n_f)}
\right]^{\frac{1}{4}}\,
\exp\left\{ \sqrt{d_0\xi\;
\tau}-
d_1\tau \right\},}
\eqno (2)$$
with
$$\xi\equiv |\log x|,\,\tau\equiv\log\dfrac{\alpha_s(Q^2_0)}{\alpha_s(Q^2)}.$$
In \eqs (1), (2) the $d_+,\,d_1,\,d_0$ are known quantities related to 
the singlet anomalous dimension. The full NLO corrections to (1), (2) may 
be found in ref. 5. There it was shown that {\it both} (1) and (2) could 
be used to give reasonable fits to the large $Q^2$ data, with 
a \chidof of the order of unity if fitting data with $x\leq 10^{-2}$ or \chidof$\sim 2$ if 
including points up to $x= 3.2\times 10^{-2}$. The hard singularity expression 
gave more satisfactory results for small $x$, but tended to disagree
 with experiment for ``large" 
values $x\gsim 10^{-2}$.

Because of this feature it was suggested in ref. 5 that one could 
perhaps conclude that $F_S$ should contain {\it both} a hard 
and a soft piece: since both (1) and (2) solve 
the QCD evolution equations, so will the sum.
 In the strict $x\rightarrow 0$ limit, (1) dominates over (2); but if 
the coefficients are such that $c_0\gg B_S$ then there will exist 
regions of relatively large values of $x$ where both contributions will be 
comparable. In fact, it was pointed out that 
a situation similar to that occurs for real photoproduction ($Q^2=0$) 
where it has been shown$^{[9,10]}$ that an expression\fonote{We have added in Eq. (3) 
the contribution of the $\rho-{\rm A}_2$ Regge trajectory, which corresponds 
in DIS language to the nonsinglet piece of the structure 
function; see below.} for the cross section
$$\sigma_{\gamma(Q^2=0)p}(s)=C_Hs^\lambda+C_P+C_{\rho}s^{-\rho},\eqno (3)$$
provides an excellent fit to the data, up to the very high HERA energies$^{[4]}$; and formulas 
similar to (3) are also known to work well in the small 
$Q^2$ region\ref{11}.

In the present note we pursue this line in two directions. First, we write 
QCD inspired phenomenological formulas that permit us to interpolate
 between the large $Q^2$ regime, with $F_2$ given by a combination of (1) 
and (2) plus a nonsinglet contribution, and the low $Q^2$ region. Here we will show 
that a natural {\it consistency} condition as $Q^2\rightarrow 0$ 
allows us to {\it determine} the 
constants $\lambda=\lambda_0,\,\rho=\rho_0$ finding values 
remarkably close to those given by the fits at {\it large} $Q^2$: 
$\lambda_0\simeq 0.470,\,\rho_0\simeq 0.522$. What is more, the 
formulas provide 
excellent fits to data, with \chidof =$\tfrac{106.2}{104-4}$ for $Q^2\leq 8.5\,\gev^2$.

Secondly, using these values $\lambda_0,\,\rho_0$, and profiting from the fact 
proved in refs. 6, 8 that these quantities are independent of $Q^2$, we fix them 
to the values found at low $Q^2$ and produce a fit to {\it high} momenta data 
with an expression\fonote{We of course 
include NLO corrections to (4) in the actual fits.}
$$\eqalign{ F_2=\langle e_q^2\rangle[F_S+F_{NS}];\kern7em \phantom{x}\cr
F_S(x,Q^2)=B_S[\alpha_s(Q^2)]^{-d_+(1+\lambda_0)}x^{-\lambda_0}\cr
+ \frac{c_0}{\xi}\left[ \frac{9\xi \tau}{4\pi^2(33-2n_f)}
\right]^{\frac{1}{4}}\,\exp\left\{\sqrt{d_0\xi\;
\tau}-
d_1\tau \right\};\cr
F_{NS}(x,Q^2)=B_{NS}[\alpha_s(Q^2)]^{-d_{NS}(1-\rho_0)}x^{\rho_0}.}
\eqno (4)$$ 
[$d_{NS}(1-\rho_0)$ is the nonsinglet anomalous dimension, a known function of $1-\rho_0$]. 
This fit improves substantially those obtained with (1) 
or (2) separately, with the addition of 
a single new parameter, since we took $\lambda_0,\,\rho_0$ given by 
the low energy calculation. We are therefore able to give 
a consistent description of data, from $Q^2=0$ to thousands of $\gev^2$, 
and from $x\sim 10^{-6}$ to $x=0.032$ with, as free 
parameters, only the constants $c_0,\,B_S,\,B_{NS}$ and $Q_0^2$, plus a low energy one 
(see below).
\vskip.4cm
\noindent\S{\bf 2. The low $Q^2$ region}. The expression for the virtual photon scattering
 cross section in terms of 
the structure function $F_2$ is
$$\sigma_{\gamma(Q^2=0)p}(s)=\dfrac{4\pi\alpha}{Q^2}F_2(x,Q^2),\;{\rm with}\; s=Q^2/x.
\eqno (5)$$
In the low energy region we should, as discussed, take the soft-Pomeron 
dominated expression (2) to be given by an ordinary Pomeron, i.e.,
 behaving as a constant for $x\rightarrow 0$ (or equivalently, $s\rightarrow \infty$). 
So the expression that will, when evolved to large $Q^2$ 
yield (4) is
$$\eqalign{F_2=\langle e_q^2\rangle\Big\{
B_S[\alpha_s(Q^2)]^{-d_+(1+\lambda_0)}x^{-\lambda_0}\cr
+C+B_{NS}[\alpha_s(Q^2)]^{-d_{NS}(1-\rho_0)}x^{\rho_0}\Big\}.}
\eqno (6)$$
Note that $C\neq c_0$ as the gluon component also intervenes
 in the evolution; the relation between $C,\, c_0$ may 
be found in refs. 8, 5, and later in the present work.

On comparing (5) and (6) we see that, as noted in ref. 10, we have 
problems if we want to extend (6) to very small $Q^2$. First of all, 
$$\alpha_s(Q^2)=\dfrac{4\pi}{\beta_0\log Q^2/\Lambdav^2}
\eqno (7)$$ {\it diverges} when $Q^2\sim\Lambdav^2$. Secondly, 
Eq. (5) contains the factor $Q^2$ in the denominator so the 
cross section blows up as $Q^2\rightarrow 0$ unless $F_2$ were to develop 
a zero there.

It turns out that there is a simple way to solve both 
difficulties at the same time. It has been conjectured$^{[12]}$ that the 
expression (7) for $\alpha_s$ should be modified for values of 
$Q^2$ near $\Lambdav^2$ in such a way that it {\it saturates}, producing in particular
 a finite value for $Q^2\sim\Lambdav^2$. To be precise, one alters (7) according to
$$\alpha_s(Q^2)\rightarrow\dfrac{4\pi}{\beta_0\log (Q^2+M^2)/\Lambdav^2},$$
where $M$ is a typical hadronic mass, $M\sim m_\rho\sim\Lambdav(n_f=2)\dots$; 
 the value $M=0.96\,\gev$ has been suggested on the basis 
of lattice calculations. Here we will simply set $M=\Lambdav=\Lambdav_{\rm eff}$, to avoid 
a proliferation of parameters. For the Pomeron term [the constant in Eq. (6)] 
we merely replace $C\rightarrow Q^2/(Q^2+\Lambdav^2_{\rm eff})$. The  
expression we will use for low $Q^2$ is thus,
$$\eqalign{F_2=\langle e_q^2\rangle\Big\{
B_S[\tilde{\alpha}_s(Q^2)]^{-d_+(1+\lambda)}Q^{-2\lambda}s^\lambda\cr
+C\dfrac{Q^2}{Q^2+\Lambdav^2_{\rm eff}}+
B_{NS}[\tilde{\alpha}_s(Q^2)]^{-d_{NS}(1-\rho)}Q^{2\rho}s^{-\rho}\Big\},}
\eqno (8{\rm a})$$
where
$$\tilde{\alpha}_s(Q^2)=\dfrac{4\pi}{\beta_0\log (Q^2+\Lambdav_{\rm eff}^2)/\Lambdav_{\rm eff}^2}
\eqno (8{\rm b})$$
and we have changed variables, $(Q^2,\,x)\rightarrow (Q^2,\,s=Q^2/x)$.

We have still not solved our problems: given Eq. (5) it is clear that 
a {\it finite} cross section for $Q^2\rightarrow 0$ will 
only be obtained if the powers of $Q^2$ match exactly. This is automatic by 
construction for the Pomeron term, but for the hard singlet and the nonsinglet 
piece it will only occur if we have consistency conditions satisfied. With 
the expression given in (8b) for $\tilde{\alpha}_s$ it diverges 
as ${\rm Const.}/Q^2$ when $Q^2\rightarrow 0$: so we only get 
a matching of zeros and divergences if $\lambda=\lambda_0,\,\rho=\rho_0$ with   
$$d_+(1+\lambda_0)=1+\lambda_0,\;d_{NS}(1-\rho_0)=1-\rho_0.\eqno (9)$$
The solution to these expressions depends very little on the number of
 flavours; for $n_f=2$, probably 
the best choice at the values of $Q^2$ we will be working with, one finds 
$\lambda_0=0.470,\,\rho_0=0.522$. The second is in uncanny 
agreement with the value obtained with either a Regge 
analysis in hadron scattering processes, or 
by fitting structure functions in DIS. The first is slightly
 larger than the value obtained in fits to DIS$^{[5,6]}$ which give  
$\lambda=0.32\;{\rm to}\;0.38$. Nevertheless, these figures 
were obtained in fits {\it omitting} the 
soft Pomeron-dominated piece; when adding it the favoured
 value (see below) is $\lambda=0.43$, and, as we will see, 
an essentially as good fit may be obtained provided $\lambda\leq 0.5$.

\setbox3=\vbox{\hsize 12.4truecm
\setbox1=\vbox{\hsize 12.4truecm \epsfxsize=12.4truecm\epsfbox{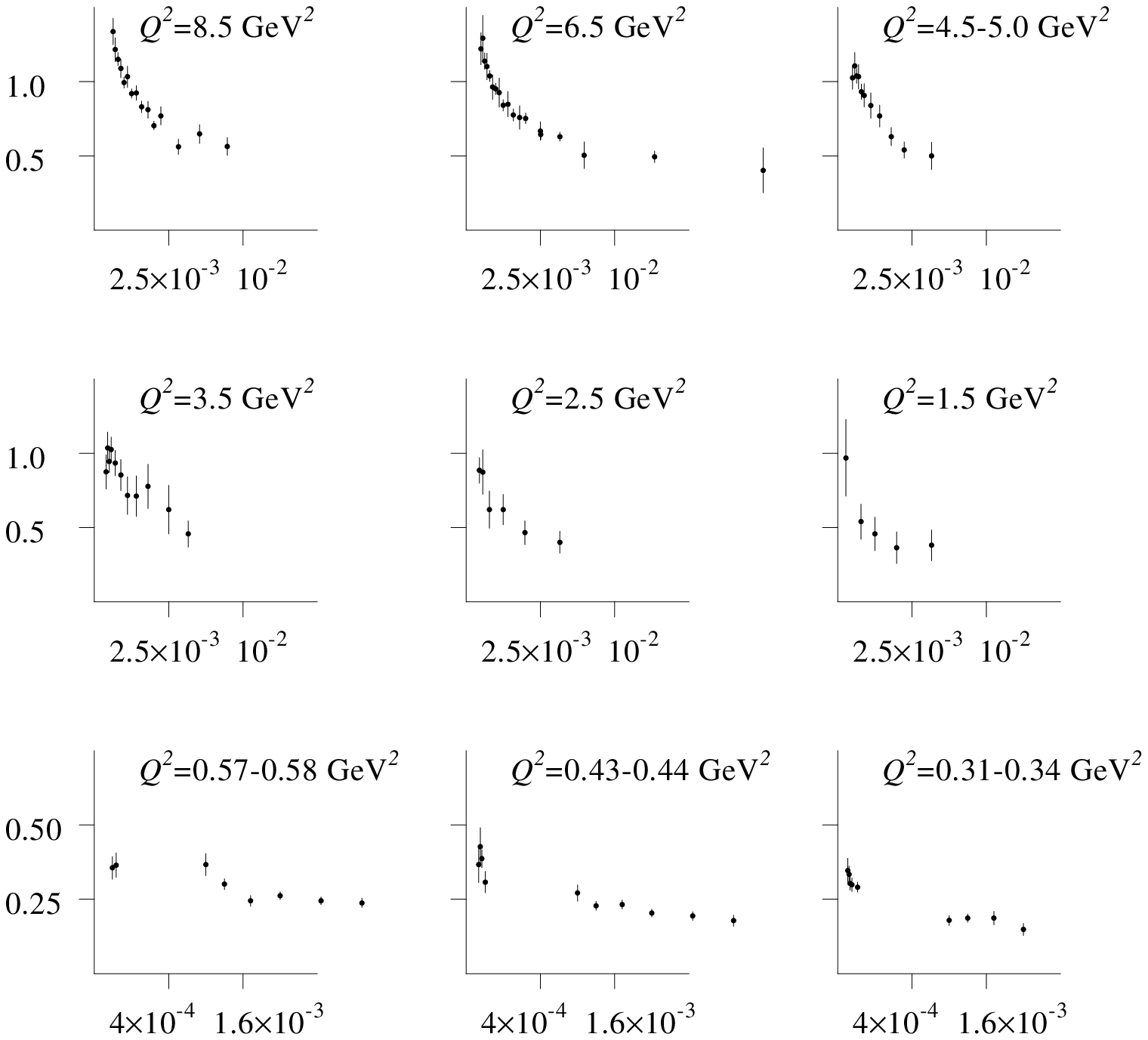}}
\setbox2=\vbox{\hsize 12.4truecm \epsfxsize=12.4truecm\epsfbox{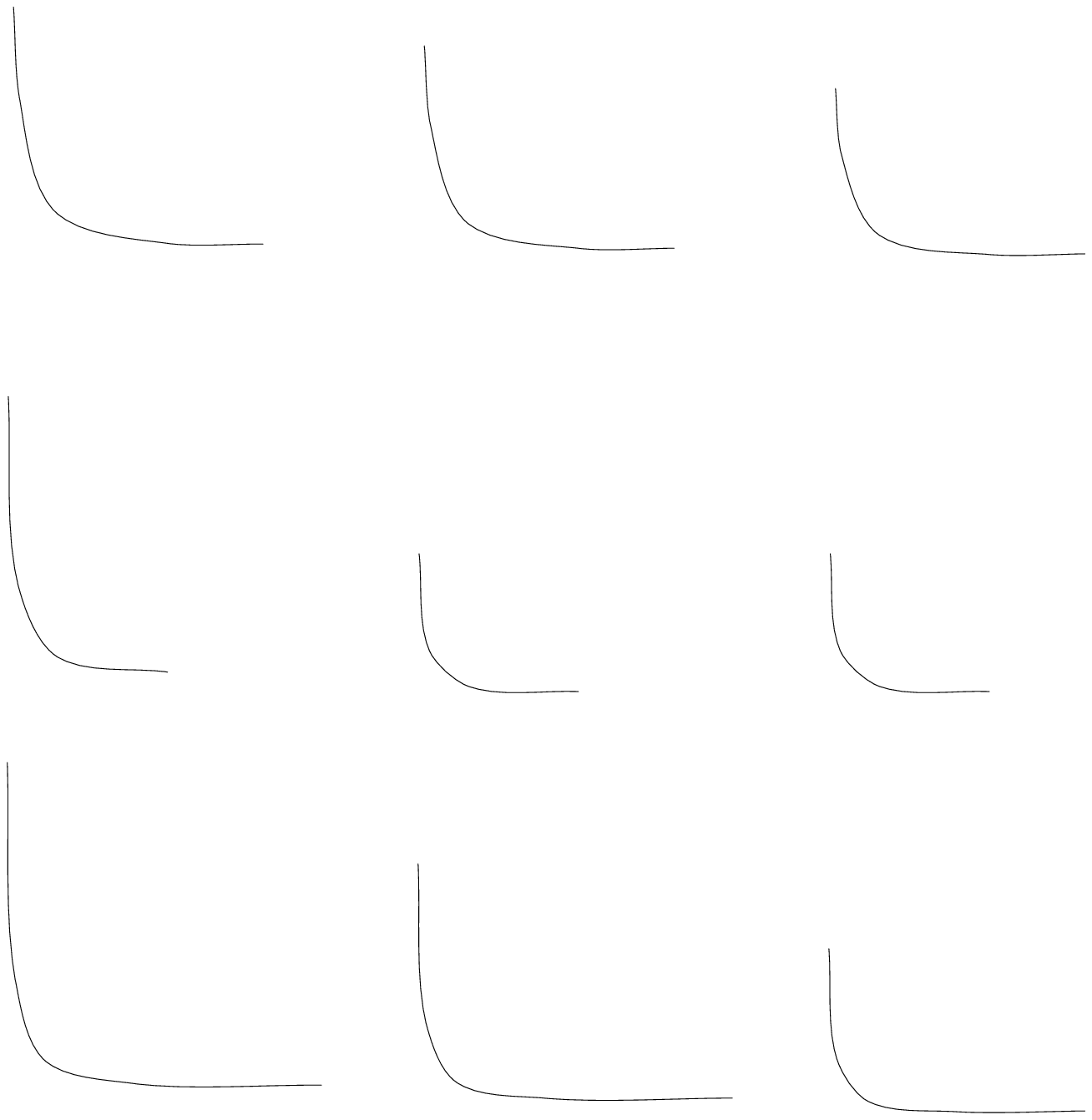}}
\centerline{\box1\kern-12.4truecm\box2}}
\setbox0=\vbox{\hsize4.1truecm \petit \noindent Figure 1. Comparison of predictions
 with data, Zeus plus H1. The
neutrino data and prediction are not shown. The 
plot is given of $F_2$ {\it vs.} $x^{1/2}$.\hb
\vskip.3cm} 
\line{\box3\hfil\box0}
\vskip.2cm

We are perfectly aware that we are pushing QCD well below its region 
of applicability, and that the condition of matching at $Q^2\rightarrow 0$ is 
at best only of phenomenological value. Nevertheless, the fact that we get 
such reasonable predictions for $\lambda_0,\,\rho_0$ probably 
indicates that our procedure represents, {\it grosso modo}, 
the actual situation, which is also justified by  
the quality of the 
fit Eq. (9) provides. If we take all H1 and Zeus data$^{[2,3]}$ for
 $0.31\,\gev^2\leq Q^2< 8.5\,\gev^2$, and we include 
10 neutrino $xF_3$ data\fonote{We choose to include these because the function $xF_3$ is 
pure nonsinglet and its presence thus helps stabilize the fits: see ref. 5 for details.} we 
find
$$\Lambdav_{\rm eff}=0.87\, \gev,\;\langle e_q^2\rangle B_S= 5.28\times10^{-3},
\;\langle e_q^2\rangle B_{NS}=0.498,\;\langle e_q^2\rangle C=0.486,$$
for a \chidof$=\tfrac{106.2}{104-4}$.
The value of $\Lambdav_{\rm eff}$ we have obtained lies somewhere inside the expected 
bracket, $\Lambdav(n_f=2)\simeq 0.35\,\gev$ and the value found in the quoted lattice 
calculation, 0.96 GeV. Clearly, the fit gives a compromise, phenomenological quantity.

 The agreement 
between phenomenology 
and experiment, shown graphically in Fig.1, is unlikely to be trivial; $x$ varies between
 $6\times 10^{-6}$ and $4\times 10^{-2}$ , and $F_2$ changes by almost one order 
of magnitude. To see this more clearly, we replace the 
hard singularity by an evolved soft Pomeron, with a saturated $\alpha_s$. That is, 
we now fit with the expression
$$\eqalign{F_2=\langle e_q^2\rangle 
\Bigg\{\frac{c_0}{\xi}
\left[ \frac{9\xi \log[\tilde{\alpha}_s(Q_0^2)/\tilde{\alpha}_s (Q^2)]}{4\pi^2(33-2n_f)}
\right]^{\frac{1}{4}}
\exp\left( \sqrt{d_0\xi\;
\left[\log\frac{\tilde{\alpha}_s(Q_0^2)}{\tilde{\alpha}_s(Q^2)}\right]}-
d_1\log\frac{\tilde{\alpha}_s(Q_0^2)}{\tilde{\alpha}_s(Q^2)} \right)
\cr
+C\dfrac{Q^2}{Q^2+\Lambdav^2_{\rm eff}}+
B_{NS}[\tilde{\alpha}_s(Q^2)]^{-d_{NS}(1-\rho)}Q^{2\rho}s^{-\rho}\Bigg\},}$$
$\tilde{\alpha}_s$ as before. Then we find
$$\Lambdav_{\rm eff}=0.41\,\gev,\;\langle e_q^2\rangle c_0=0.094,\;\langle e_q^2\rangle C=0.253,
\;\langle e_q^2\rangle B_{NS}=0.41$$
and a \chidof =250/(104-4). We consider these results as convincing proof of the 
necessity of a hard component, which will be henceforth assumed.
\vskip.4cm
\noindent\S{\bf 3. The high $Q^2$ region}. We next discuss the large $Q^2\gsim 10\,\gev^2$ 
region, under various hypotheses for the small $Q^2$ region, which we then evolve with QCD. 
We will consider moderately large values of $x$,  
 $x\leq 0.032$ because we will be interested not 
only on the leading behaviour as $x\rightarrow 0$, given almost certainly 
by a hard singularity, but also on the subleading corrections. We will, because 
of the success of the small $Q^2$ fits just discussed, assume for the leading piece 
a hard Pomeron term with a critical value for $\lambda=\lambda_0$. This will be 
also justified a posteriori both by the quality of the fits and by the fact that 
if, in the best alternative (which will 
turn out to be a hard plus a soft Pomeron), we leave $\lambda$ as a free
 parameter, its optimum falls very close 
to $\lambda_0$; see below. So we assume that at a certain, fixed $Q^2_0\sim 1\,\gev^2$, one has
$$F_S(x,Q^2_0)\simeq B_Sx^{-\lambda_0}+F_{\rm corr.}(x,Q^2_0).$$
For the correction term, $F_{\rm corr.}(x,Q^2_0)$, we consider the following 
possibilities: a soft Pomeron,
$$F^P_{\rm corr.}(x,Q^2_0)\simeq {\rm constant};\eqno (10{\rm a})$$
and a $P'$ Regge pole,\fonote{For Regge pole theory, and the triple (and multiple) Pomeron 
terms cf. respectively refs. 13, 14. }
$$F^{P'}_{\rm corr.}(x,Q^2_0)\simeq {\rm constant}\times x^{1-\alpha_{P'}(0)},
\alpha_{P'}(0)\sim 0.5 .\eqno (10{\rm b})$$
We also consider the possibility of adding a triple Pomeron-induced term,
$$F^{TP}(x,Q^2_0)\simeq {\rm constant}\times x^{-2\lambda_0}.\eqno(10{\rm c})$$

Once assumed the behaviours given in (10), and taking 
for simplicity that the gluon
 structure function behaves like the quark singlet one, we 
evolve with QCD for higher $Q^2$. Following the methods of refs. 5, 6, 8 
we find, to NLO,
$$F_S(x,Q^2)\simeq 
B_S\left\{1+\dfrac{c_S(1+\lambda_0)\alpha_s}{4\pi}\right\}
\ee^{q_S(1+\lambda_0)\alpha_s/4\pi}
 [\alpha_s(Q^2)]^{-d_+(1+\lambda_0)}x^{-\lambda_0}+F_{\rm corr.}(x,Q^2),\eqno (11)$$
and the values of the quantities $d_+,\,c_S,\,q_S$ may be found in ref. 5. Then, and 
depending on the low $Q^2_0$ hypothesis we make we find the correction terms,
$$\eqalign{F^P_{\rm corr.}(x,Q^2_0)\simeq \left\{1+2\sqrt{\dfrac{\xi}{ d_0\tau}}\left[k_1\;
\dfrac{\alpha_s(Q^2)}{4\pi}-
k\;\dfrac{\alpha_s(Q^2_0)}{4\pi}\right]\right\}\cr
\times \frac{c_0}{\xi}\left[ \frac{9\xi \tau}{4\pi^2(33-2n_f)}
\right]^{\frac{1}{4}}
\exp\left\{ \sqrt{d_0\xi\;
\tau}-
d_1\tau \right\},}\eqno (12{\rm a})$$
$k,\,k_1$ also given in ref. 5. For the $P'$ pole we find a very 
similar formula:
$$\eqalign{F^{P'}_{\rm corr.}(x,Q^2_0)\simeq 
\left\{1+2\sqrt{\dfrac{\xi}{ d_0\tau}}\left[k_1\;
\dfrac{\alpha_s(Q^2)}{4\pi}-
k\;\dfrac{\alpha_s(Q^2_0)}{4\pi}\right]\right\}\cr
\times\dfrac{ c_{P'}}{\sqrt{\xi}}\left(\dfrac{\tau}{\xi}\right)^{\frac{3}{4}}\exp\left\{\sqrt{d_0\xi\;
\tau}-
d_1\tau \right\}.}
\eqno (12{\rm b})$$
Finally, for the triple Pomeron-induced term,
$$F^{TP}(x,Q^2_0)\simeq  
B_{TP}\left\{1+\dfrac{c_S(1+2\lambda_0)\alpha_s}{4\pi}\right\}
\ee^{q_S(1+2\lambda_0)\alpha_s/4\pi}
 [\alpha_s(Q^2)]^{-d_+(1+2\lambda_0)}x^{-2\lambda_0}.\eqno (12{\rm c})$$ 
 
As we will see, none of the three possibilities gives a really good fit 
in the ``large" $0.01<x\leq 0.032$ region; for the more 
precise Zeus data the \chidof  is of 1.7. To remedy this we consider the 
possibility of softening the large $x$ region by multiplying $F_S(x,Q^2)$ by a factor 
$(1-x)^{\nu}$: for the discussion and justification of this 
we send to refs. 5, 10.

The results are summarized in Table I, where we compare the fits obtained with (12a) with 
the fits found in ref. 5 using only the hard singularity or soft Pomeron-dominated 
expressions, i.e., setting respectively $c_0,\,B_S$ equal to zero.
\vskip.2cm
\setbox0=\vbox{\hsize=15cm
{\petit 
\vskip.2cm
\centerline{{\bf Table ${\rm I}a$}.- $n_f=4$; H1 data. $x\leq 0.032,\,Q^2\geq 12\,\gev^2$}
\centerrule{16em}
\vskip.1cm
 $$\matrix{
\matrix{{\rm Hard \;singularity,}\cr{\rm large\;}x\;{\rm softened}}\left\{
\matrix{\Lambdav & \lambda & \langle e^2_q\rangle  B_S & \langle e^2_q\rangle  B_{NS}&\chi^2/{\rm d.o.f.}\cr
0.140\,\gev&0.318&1.78\times10^{-4}&0.35&\tfrac{134}{110-4}}\right.\cr
&\phantom{x} & & &\cr
{\rm Soft\;Pomeron\;only}\left\{
\matrix{\Lambdav & Q_0^2& \langle e^2_q\rangle c_0 & \langle e^2_q\rangle  B_{NS}&\chi^2/{\rm d.o.f.}\cr
0.165\;\gev&0.70\,\gev^2&0.265&0.246&\tfrac{191}{110-4}}\right.\cr
&\phantom{x} & & & &\cr
{\rm Hard+Soft\;Pomeron^*}\left\{
\matrix{\lambda\;({\rm fixed})& Q_0^2& \langle e^2_q\rangle c_0 &
\langle e^2_q\rangle B_S& \langle e^2_q\rangle  B_{NS}&\chi^2/{\rm d.o.f.}\cr
0.47&1.75\;\gev^2&0.226&4.33\times10^{-4}&0.288&\tfrac{138}{110-4}}\right.
\cr
&\phantom{x} & & & &\cr
\matrix{{\rm Hard+Soft,}\cr{\rm large\;}x\;{\rm softened^*}}\left\{
\matrix{\lambda\;({\rm fixed})& Q_0^2& \langle e^2_q\rangle c_0 &
\langle e^2_q\rangle B_S& \langle e^2_q\rangle  B_{NS}&\chi^2/{\rm d.o.f.}\cr
0.47&2.22\;\gev^2&0.397&3.36\times10^{-4}&0.353&\tfrac{65.7}{110-4}}\right.}
$$
\vskip.1cm
$^*\;\Lambdav\;{\rm fixed\;at}\;0.230\;\gev.$}
}
\centerline{\boxit{\box0}}
\vskip.2cm
For the Zeus data one gets similar results, with somewhat less good \chidof .
\vskip.2cm
\setbox0=\vbox{\hsize=15.cm
{\petit 
\vskip.2cm
\centerline{{\bf Table ${\rm I}b$}.- $n_f=4$; Zeus data. $x\leq 0.025,\,Q^2\geq 12.5\,\gev^2$}
\centerrule{16em}
\vskip.1cm
 $$\matrix{
\matrix{{\rm Hard \;singularity},\cr 
{\rm only\;}x\leq 0.01\;{\rm data}}\left\{
\matrix{\Lambdav & \lambda & \langle e^2_q\rangle  B_S & \langle e^2_q\rangle  B_{NS}&\chi^2/{\rm d.o.f.}\cr
0.135\,\gev&0.301&1.25\times10^{-4}&0.314&\tfrac{126}{92-4}}\right.\cr
&\phantom{x} & & & &\cr
{\rm Soft\;Pomeron\;only}\left\{
\matrix{\Lambdav & Q_0^2& \langle e^2_q\rangle c_0 & \langle e^2_q\rangle  B_{NS}&\chi^2/{\rm d.o.f.}\cr
0.165\;\gev&0.90\,\gev^2&0.282&0.240&\tfrac{273}{120-4}}\right.\cr
&\phantom{x} & & & &\cr
{\rm Hard+Soft\;Pomeron^*}\left\{
\matrix{\lambda\;({\rm fixed})& Q_0^2& \langle e^2_q\rangle c_0 &
\langle e^2_q\rangle B_S& \langle e^2_q\rangle  B_{NS}&\chi^2/{\rm d.o.f.}\cr
0.47&2.45\;\gev^2&0.252&4.42\times10^{-4}&0.294&\tfrac{197}{120-4}}\right.
\cr
&\phantom{x} & & & &\cr
\matrix{{\rm Hard+Soft,}\cr 
{\rm large\;}x\;{\rm softened^*}}\left\{
\matrix{\lambda\;({\rm fixed})& Q_0^2& \langle e^2_q\rangle c_0 &
\langle e^2_q\rangle B_S& \langle e^2_q\rangle  B_{NS}&\chi^2/{\rm d.o.f.}\cr
0.47&2.72\;\gev^2&0.310&3.417\times10^{-4}&0.370&\tfrac{143}{120-4}}\right.}
$$
\vskip.1cm
$^*\;\Lambdav\;{\rm fixed\;at}\;0.230\;\gev.$
 The optimum value would correspond to $\Lambdav\sim 0.45\;\gev$.}
}
\centerline{\boxit{\box0}}            
\vskip.2cm
In both tables the expression ``large $x$ softened" means that we have multiplied the 
formulas for $F_S$ by a factor $(1-x)^\nu$, to correct the structure functions for (relatively) 
large values of $x$. For the hard singularity case, cf. ref. 5; for the Hard + Soft singularities 
cases, we have taken $\nu=10\sim 11$. It is also to be remarked that, if we had {\it fitted} $\lambda$ 
to e.g., the Zeus data using a hard plus soft term, we would 
have obtained (not correcting for large $x$, and fixing $\Lambdav=0.23\,\gev$),
$$\lambda=0.429,\;Q_0^2=2.40\;\gev^2,\; \langle e^2_q\rangle c_0=0.244,
\; \langle e^2_q\rangle B_S=3.83\times10^{-4},\; \langle e^2_q\rangle B_{NS}=0.30,$$
for a \chidof $=\tfrac{195}{120-5}$. Clearly the values of the parameters, and the \chidof  
are so similar to those obtained in Table I{\it b} fixing $\lambda$ to its 
low energy value found before,\fonote{It is of some interest 
to remark that the optimum $\lambda$ found is actually practically identical 
to the critical value we would have obtained at low $Q^2$ with {\it zero} flavours, $\lambda_{n_f=0}=0.447$. 
The difference with the value for two flavours, and the uncertainties inherent to the 
analysis, however, make us stick 
to the choice $\lambda_{n_f=2}=0.470$.} $\lambda=0.47$, that 
we feel justified {\it a posteriori} in taking such value also at large $Q^2$. 

As next possibility, we try a hard Pomeron, plus a $P'$ Regge trajectory, Eq. (12b). If we fit 
the Zeus data with this, adding a softening factor $(1-x)^\nu$, $\nu=11$, we find the results of 
Table II.
\vskip.2cm
\setbox0=\vbox{\hsize=14.cm
{\petit 
\vskip.2cm
\centerline{{\bf Table {\rm II}}.- $n_f=4$; Zeus data.}
\centerrule{16em}
\vskip.1cm
 $$\matrix{{\rm Hard}+P',\cr
 {\rm large\;}x\;{\rm softened^*}} \left\{
\matrix{\lambda\;({\rm fixed})& Q_0^2& \langle e^2_q\rangle c_{P'} &
\langle e^2_q\rangle B_S& \langle e^2_q\rangle  B_{NS}&\chi^2/{\rm d.o.f.}\cr
0.47&1.11\;\gev^2&0.616&4.25\times10^{-4}&0.343&\tfrac{171}{120-4}}\right.
$$
\vskip.1cm
$^*\;\Lambdav\;{\rm fixed\;at}\;0.230\;\gev.$}
}
\centerline{\boxit{\box0}}            
\vskip.2cm
\noindent If we had not corrected for the large $x$ values, i.e., we had not included the 
factor $(1-x)^\nu$, we would have obtained a \chidof  of 270.
 
We finish this section by describing the 
results of the fits including a triple Pomeron-induced term. That is, we 
consider that 
$$F_S=F_H+F_{\rm corr.}+F^{TP}.\eqno (13)$$
Then we get, for the Zeus data,
\vskip.2cm
\setbox0=\vbox{\hsize=14.cm
{\petit 
\vskip.2cm
\centerline{{\bf Table {\rm III}}.- $n_f=4$; Zeus data. Triple Pomeron correction added.}
\centerrule{16em}
\vskip.1cm
 $$\matrix{{\rm Hard}+P\cr
 +{\rm TP\; term^*}} \left\{
\matrix{\lambda\;({\rm fixed})& Q_0^2& \langle e^2_q\rangle c_{P} &
\langle e^2_q\rangle B_S& \langle e^2_q\rangle B_{TP}& \langle e^2_q\rangle  B_{NS}&
\chi^2/{\rm d.o.f.}\cr
0.47&3.35\,\gev^2&0.201&8.67\times10^{-4}&-1.96\times10^{-4}&0.315&\tfrac{185}{120-5}}\right.
$$
\vskip.1cm
$^*\;\Lambdav\;{\rm fixed\;at}\;0.230\;\gev.$}
}
\centerline{\boxit{\box0}}            
\vskip.2cm
For H1 we would have obtained similar values for the parameters and a \chidof =105/(110-5). 
The inclusion of the triple Pomeron term improves slightly the fit; but perhaps more 
interesting is that the coefficient of it is, as expected from 
multi-Pomeron theory,\ref{14} small (when compared with the hard Pomeron) and negative. 

Clearly, the best fit is obtained with the hard plus soft Pomerons, just as for the 
low $Q^2$ region. Not only the \chidof 
is quite good, but the values of the parameters are 
very reasonable. In particular, the consistency of the picture is shown by the fact that the 
value of $Q^2_0$ where $F^P_{\rm corr.}$ is 
supposed to behave like a constant falls in the middle of the low $Q^2$ region 
where we showed that precisely this assumptions leads to a good fit to data.
 Because of this, we present in Table IV the parameters of 
the fits to, simultaneously, H1 and Zeus data on $ep$, plus neutrino data. This gives 
our best set of formulas, providing an excellent
 fit to experiment in a very wide range of $Q^2,\,x$.
\vskip.2cm
\setbox0=\vbox{\hsize=14.cm
{\petit 
\vskip.2cm
\centerline{{\bf Table ${\rm IV}a$}.- $n_f=4$; Zeus plus H1 data; $Q^2\geq 10\,\gev^2,\,x\leq 0.01$.}
\centerrule{16em}
\vskip.1cm
 $$\matrix{{\rm Hard}+P \left\{
\matrix{\lambda\;({\rm fixed})& Q_0^2& \langle e^2_q\rangle c_{P} &
\langle e^2_q\rangle B_S& \langle e^2_q\rangle  B_{NS}&
\chi^2/{\rm d.o.f.}\cr
0.47&2.95\,\gev^2&0.296&4.28\times10^{-4}&0.349&\tfrac{138.3}{144-4}}\right.\cr
\matrix{{\rm Hard}+P,\cr
 +{\rm TP\; term}} \left\{\matrix{\lambda\;({\rm fixed})& Q_0^2& \langle e^2_q\rangle c_{P} &
\langle e^2_q\rangle B_S& \langle e^2_q\rangle B_{TP}& \langle e^2_q\rangle  B_{NS}&
\chi^2/{\rm d.o.f.}\cr
0.47&4.45\,\gev^2&0.258&8.33\times10^{-4}&-1.67\times10^{-4}&0.359&\tfrac{129.3}{144-5}}\right.}
$$
\vskip.1cm
$\Lambdav\;{\rm fixed\;at}\;0.230\;\gev.$}
}
\centerline{\boxit{\box0}}            
\vskip.2cm
\setbox0=\vbox{\hsize=14.cm
{\petit 
\vskip.2cm
\centerline{{\bf Table ${\rm IV}b$}.- $n_f=4$; Zeus plus H1 data;
 $Q^2\geq 10\,\gev^2,\,x\leq 0.032$. $x$ ``softened".}
\centerrule{16em}
\vskip.1cm
 $${\rm Hard}+P \left\{
\matrix{\lambda\;({\rm fixed})& Q_0^2& \langle e^2_q\rangle c_{P} &
\langle e^2_q\rangle B_S& \langle e^2_q\rangle  B_{NS}&
\chi^2/{\rm d.o.f.}\cr
0.47&2.28\,\gev^2&0.311&2.72\times10^{-4}&0.315&\tfrac{227.4}{230-5}}\right.
$$
\vskip.1cm
$\Lambdav\;{\rm fixed\;at}\;0.230\;\gev.$}
}
\centerline{\boxit{\box0}}            
\vskip.2cm

In the second case (Table IV$b$) we do not give the fit including a triple Pomeron term 
as the \chidof  does not vary appreciably if including it provided 
$ \langle e^2_q\rangle |B_{TP}|\lsim 2\times 10^{-4}$. We consider
 the parameters given in Table IV$a$ 
to be the more reliable ones for describing low $x$ structure functions. If we had 
fitted also $\lambda$ with the whole set of data we would have obtained minima 
for values comprised between 0.42 and 0.49, with a variation of 
the chi-squared of less than two units with respect to the one obtained 
fixing $\lambda=0.470$. Finally, if we fit the QCD 
parameter $\Lambdav$, the values 
 which provide minima vary between 0.555 \gev$\,$  and 0.310 \gev, and the chi-squared 
improves by less than five units. Because of this we 
consider, as stated, that it is justified to favour the 
fits obtained with {\it fixed} $\lambda=0.470,\,\Lambdav=0.23\,\gev$.

\setbox0=\vbox{\hsize 13.5truecm \epsfxsize=13.4truecm\epsfbox{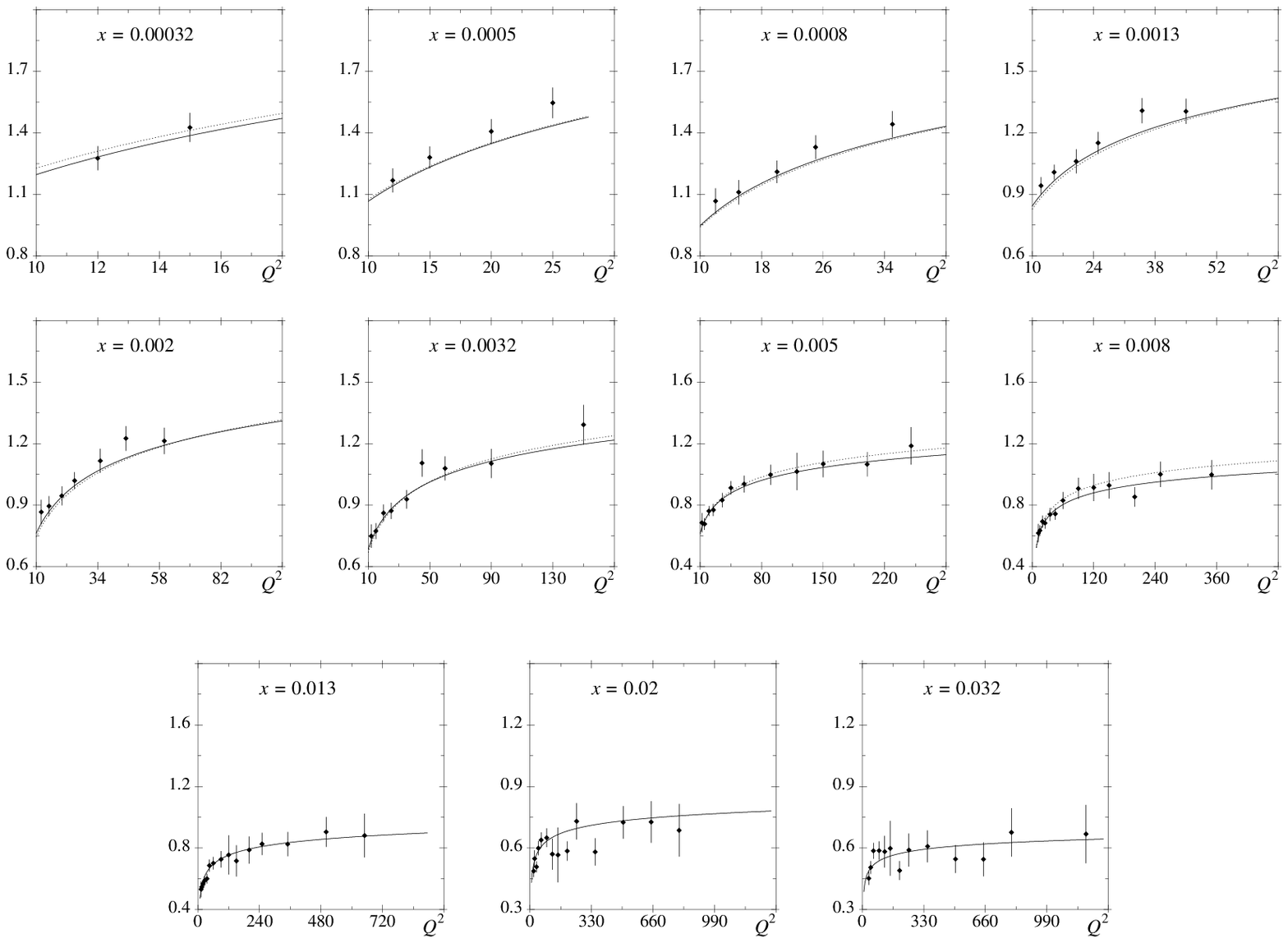}\hb{\petit
\noindent Figure 2a. Comparison of predictions from 
 Table IV with H1 $ep$ data$^{[2]}$ for $F_2$.
\vskip.1cm}}
\centerline{\box0}
\setbox0=\vbox{\hsize 13.5truecm \epsfxsize=13.4truecm\epsfbox{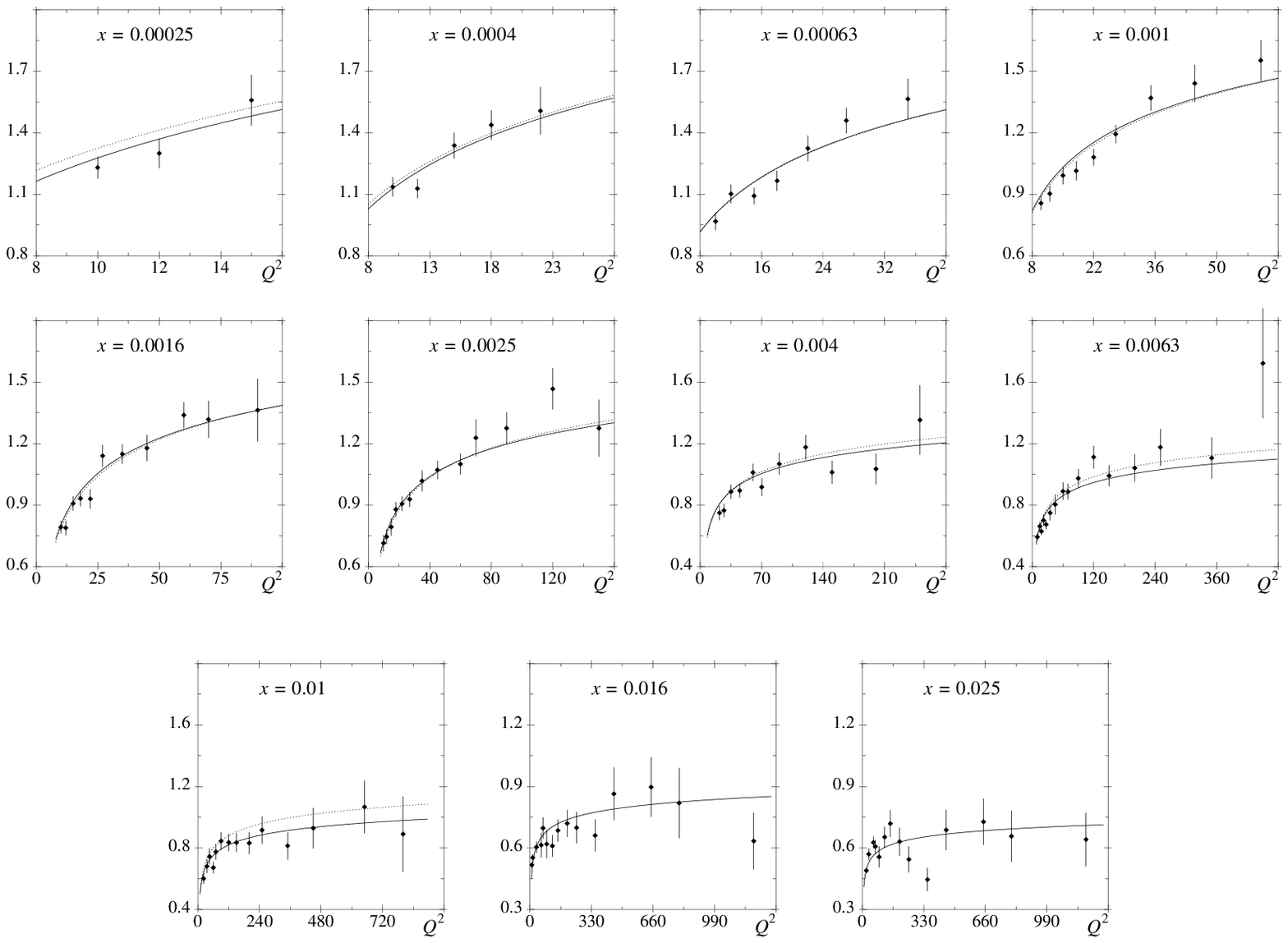}\hb{\petit
\noindent Figure 2b. Comparison of predictions from 
 Table IV with Zeus $ep$ data$^{[3]}$ for $F_2$.
\vskip.2cm}}
\centerline{\box0}

In Figs. 2 we show the comparison of our fits, with the parameters in Table IV, with data. Note 
that the {\it same} values of the parameters are used in  Fig. 2a  and Fig. 2b.
 For both Figs. 2 we give the fits with the 
``softened" and straight formulas:
 the  continuous lines indicate large-$x$ softening, and the dotted 
lines no softening.
\vskip.4cm
\noindent\S{\bf 4. Gluon and longitudinal structure functions.} Detailed predictions
 for the gluon and longitudinal structure functions are obtained trivially 
by adding the soft and hard Pomeron expressions given in ref. 5. We 
will leave the details of this to the reader; but 
we would like to comment here on a particularly interesting prediction of
 our analysis for the growth of  
 the cross-section $\sigma_{\gamma p\rightarrow J/\psi p}(W)$ as 
a function of the c.m. energy, $W$. In fact, this cross section may be expressed 
as a function of the gluon structure function $F_G$, 
$$\sigma_{\gamma p\rightarrow J/\psi p}(W)=AF_G(\bar{x}=
a\dfrac{M_{J/\psi}^2}{W^2}, Q^2=M_{J/\psi}^2),$$
and $A,\,a$ are constants approximately known. 
In the case in which the dominant singularity is a hard one, one can obtain,
 parameter-free 
the gluon structure function as\ref{6}
$$F_G(x,Q^2)\simeqsub_{x\rightarrow 0} B_G[\alpha_s]^{-d_+}x^{-\lambda_0}$$
and $B_G$ may be calculated in terms of $\lambda_0,\,B_S$. So, using above 
formulas we have, for the logarithmic 
{\it slope} of the cross section,
$$\delta\equiv\dfrac{\log\sigma_{\gamma p\rightarrow J/\psi p}(W)}{\log W}\rightarrow 2\lambda_0=0.94,$$
The numerical value is obtained with the $\lambda_0$ found in the previous analyses. This 
is in remarkable agreement with the figure reported in a
 fit\ref{15} including recent HERA data\ref{16} which gives $\delta=0.9$. A more detailed 
calculation would produce the whole of $F_G$ including corrections 
subleading as $x\rightarrow 0$; this unfortunately involves a new  
 unknown constant, the soft Pomeron component of the gluon. 
\vskip.4cm
\noindent\S{\bf 5. Discussion.} The main outcome of our 
analysis in the present note is that we are able to give a unified,
 consistent description of small $x$ DIS data, both for large and small values of $Q^2$. 
Besides this, there are a number of specific points to which we would
 like to draw also attention.

 First of all there is the matter of the dependence of 
our low $Q^2$ results on the saturation hypothesis for $\alpha_s$. It is clear 
that the good quality of the fits indicates that, 
with suitable modifications, QCD may give a {\it phenomenological} 
description of the data down to very low momenta; but of course this should not be construed 
as a proof of saturation, in particular of the very specific form considered 
here. One may interpret our results, however, as showing that the saturation expression 
is particularly adapted to represent, in DIS, a variety of effects: higher 
twists, renormalons, and likely also genuine saturation.

A second question is the connection between low and high $Q^2$. For the hard 
piece there is no problem, as both expressions are identical up to 
NLO corrections. For the soft piece, if we 
start with a constant behaviour for $Q^2\sim 2\,-\,5\,\gev^2$, then as $Q^2$ 
grows an expression like (2) will start to develop. the details of 
this will depend on what one assumes for the gluon structure function. 
Because the variation both with $Q^2$ and $x$ of the soft piece is slower 
than that of the hard part, we think the best procedure 
is to assume constancy of the soft piece up to $Q^2=8.5\,\gev^2$, and 
the evolved form from there on; since a very good fit is obtained at the low 
momentum region already with the constant behaviour there is little point in adding frills, and a new 
constant (the soft component of the gluon structure 
function).

Finally we devote a few words to the matter of multi-Pomeron exchange. If, at 
a fixed $Q^2$ a single hard Pomeron gives\fonote{The following discussion is rather 
sketchy; details and references may be found in the review of ref. 14.}
$$F_{1P}(x,Q^2)\simeqsub_{s\rightarrow\infty}b_{1P}(Q^2)s^{\lambda},$$
then an $n$-Pomeron term will produce the behaviour 
$$F_{nP}(x,Q^2)\simeqsub_{s\rightarrow\infty}b_{nP}(Q^2)s^{n\lambda};$$
the constants $b_{nP}$ should depend on the momentum at 
which they are calculated.

In some approximations (e.g., of eikonal type\ref{14}) one has, for $Q^2=-M^2_{\rm had}$, 
i.e., for on-shell scattering of hadrons,
$$b_{nP}(-M^2_{\rm had})=(-1)^{n+1}\dfrac{\delta^n}{n!}C,$$
so we get for the sum
$$\eqalign{F_S(x,-M^2_{\rm had})=\sum_{n=1}^{\infty}F_{nP}(x,Q^2=-M^2_{\rm had})\cr
=C-C\exp\left[-\delta s^{\lambda}\right]\simeqsub_{s\rightarrow\infty}C.}$$
For values of $Q^2$ of the order of $Q^2_0\sim 2\,-\,4\,\gev^2$, we expect that 
the $b_{nP}(Q^2)$ will not change much, so if we write
$$b_{nP}(Q_0^2)\simeq b_{nP}(-M^2_{\rm had})+\Deltav_n.$$
we will then get,
$$F_S(x,Q_0^2)\simeq F_S(x,-M^2_{\rm had})+\sum \Deltav_n\simeq C+\Deltav_1 x^{-\lambda}+
\Deltav_2 x^{-2\lambda}+\dots$$
which is the expression we have used in the text. In this respect, the evidence for 
a behaviour like this may be taken as supporting multi Regge theory.  

\vfill\eject
\noindent{\bf ACKNOWLEDGEMENTS}
\vskip.3cm
The financial help of CICYT, Spain, is gratefully acknowledged. Thanks are due to F. Barreiro and 
A. Kaidalov for illuminating discussions.
\vskip.4cm
\centerline{\bf REFERENCES}
\vskip.3cm
{\petit
\noindent 1. M. Derrick et al, Z. Phys. {\bf C65} (1995) 397; Phys. Lett. {\bf B345} 
(1995) 576.\hb
2. H1 Collaboration, preprint DESY 96-039, 1996.\hb 
3. Zeus Collaboration, preprint DESY 96-076, 1996.\hb
4. M. Derrick et al., \jpl{295}{1992}{465}; {\it ibid.}, \jzp{63}{1994}{399}.\hb 
5. K. Adel, F. Barreiro and F. J. Yndur\'ain, FTUAM 96-39 (hep-hp/9610380), in press in Z. Phys. C.\hb 
6.  C. L\'opez and F. J. Yndur\'ain, Nucl. Phys., {\bf B171} (1980) 231.\hb 
7. A. De R\'ujula et al., Phys. Rev., {\bf D10} (1974) 1649.\hb 
8. F. Martin, Phys. Rev., {\bf D19} (1979) 1382. \hb 
9. C. L\'opez and F. J. Yndur\'ain, Phys. Rev. Lett., {\bf 44} (1980) 1118.\hb 
10. F. J. Yndur\'ain, Preprint FTUAM 96-12 (revised),
 to be published in Proc. QCD 96, Nucl. Phys. Suppl.\hb
11. B. Badelek and J. Kwiecinski, Phys. Lett. {\bf B295} (1992) 263; Rev. Mod. Phys.,
 {\bf 68} (1996) 445.\hb
12. Yu. A. Simonov, Yadernaya Fizika, {\bf 58} (1995) 113, and work quoted there.\hb
13. V. D. Barger and D. B. Cline, {\sl Phenomenological Theories 
of High Energy Scattering}, Benjamin, 1969.\hb
14. A. B. Kaidalov, Survey in High Energy Physics, Vol. 9 (1996) 143.\hb
15. H. Klein, DESY 96-218 (1996).\hb
16. M. Derrick et al., \jpl{350}{1995}{134}, and work quoted there.\hb  
}  
\bye